\documentclass[10pt,final,journal,finalsubmission,cspaper,compsoc]{IEEEtran}

\usepackage[pdftex]{graphicx}

\usepackage[cmex10]{amsmath}
\usepackage{mathtools}

\usepackage{enumerate}

\ifCLASSOPTIONcompsoc
  \usepackage[caption=false,font=normalsize,labelfont=sf,textfont=sf]{subfig}
\else
  \usepackage[caption=false,font=footnotesize]{subfig}
\fi

\begin{document}
\title{A Design Methodology for \\Software Measurement Programs}

\author{Alejandro~S\'anchez~Guinea
\IEEEcompsocitemizethanks{\IEEEcompsocthanksitem A. Sanchez Guinea is with the 
Computer Science Department,\protect\\ University of Helsinki, Finland.\protect\\
E-mail: azsanche@cs.helsinki.fi}}

\markboth{IEEE TRANSACTIONS ON SOFTWARE ENGINEERING,~Vol.~X, No.~X, January~XXXX}%
{S\'anchez Guinea: A Design Methodology for Software Measurement Programs}

\IEEEcompsoctitleabstractindextext{%
\begin{abstract}

Software measurement programs have emerged as compounds of several
measurement activities that are pursued as part of a combined effort
of several parties within a software organization, based on interests 
that the organization has regarding the assessment of the different elements
that intervene in the development of software.

This paper recognizes design of measurement programs as
an essential activity that, up until now, has been studied extensively, however,
only in what respects to the content of the programs. In addition, proper
specification for this kind of programs, accounting for preciseness and unambiguity,
to facilitate maintenance, evolution, and execution has not been thoroughly considered.
A methodology for designing programs that embody these and some other desirable features
is presented. The methodology is built in solid ground. From software measurement literature,
a goal-oriented approach is considered for building the content of the program. On the other hand,
a successful technique from software development as modularization is utilized
to give coherent structure to the measurement program.

\end{abstract}

}

\maketitle

\IEEEdisplaynotcompsoctitleabstractindextext

\IEEEpeerreviewmaketitle

\section{Introduction}

\IEEEPARstart{A}{n} engineering discipline in order to be complete,
needs to define mechanisms for measuring the objects it
produces, and the processes and devices it uses for
such production. 
At this respect, measurement has been studied for many years within
Software Engineering, focusing on different points of
interest. Measurement principles have been introduced
in the context of software \cite{basili-rombach:88, fenton:94}, 
different mechanisms
have been described for the specification of the processes
involved in measuring software \cite{morisio:99}, and
 methodologies for defining
new metrics with enough foundation and mathematical rigor
have been developed \cite{kitch:95, briand:96, briand:02}.
Consequently, the combination of all such points of interest derived into
the study of full measurement endeavors, recognized as software measurement
programs \cite{gopal:02}, or as software metrics programs \cite{plfeeg:93}.

For measurement programs, the top-down design has been widely accepted as the 
line to be followed. Several works, coming mainly from the domain 
of Empirical Software Engineering have supported this type of design,
better known as goal-oriented approach 
\cite{basili-weiss:84, basili-rombach:88, briand-practical:96}.
This approach represents a comprehensive and systematic
way of eliciting measurement goals that derive into the metrics
that will drive the measurement activities over the different objects of 
measurement involved in the development of software.

The main representative of the goal-oriented approach is the
Goal-Question-Metric (GQM) method, originally introduced by
Basili and Weiss \cite{basili-weiss:84}. GQM evolved first to
include templates that provide support on the elicitation of 
measurement goals, by breaking them down into
components such as purpose, focus, object of measurement,
viewpoint, and context \cite{basili-rombach:88}.
After that, the method was included as part of
a paradigm for the improvement of software organizations 
denominated Quality Improvement Paradigm (QIP) \cite{basili:92, basili:93},
that further extended the measurement goal templates
into documents called abstract sheets \cite{briand-practical:96}, considering aspects
related to the elicitation of goals such as
quality focus, baseline hypothesis, and variation factors. 

Recently, some extensions of the GQM approach have been proposed in
an attempt to involve in the design of measurement programs, more and in a better way,
the high-level goals
of the organization that is pursuing the measurement endeavor 
\cite{offen:97,becker:99, alejandro:01, basili:07_1}. 
From
these ones, GQM${}^+$Strategies \cite{basili:07_1, basili:07, basili:10} for instance, 
makes explicit the linkages between
goals at different levels of the program, from business level, passing
through software level, until goals that reside at the products and processes level. 
In addition, GQM${}^+$Strategies includes specialized templates
for organizational goals, and considers the existent relations between different goals, focusing
in making explicit potential conflicts between them.

All approaches mentioned above, focus on producing the content of
a software measurement program, leaving out of consideration 
essential aspects such as the structure of the program, its data and 
control flow, the precise specification of the relations
between goals at same and different levels of concerns, 
as well as the dependencies between metrics all along the program. 
Furthermore, these approaches treat a measurement program
 as a single block that comprises
all measurement processes, being in this way, opposite to the general view that is
common and highly successful among engineering disciplines, where a system or a program,
depending on the case, it is broken down into parts in order to facilitate
all phases of its design and development \cite{baldwin:00}.

Within Software Engineering, it has been proposed for the
design of computer programs, to divide the structure of a program
into partitions or modules that may hold connections among them,
in which each module represents a subprogram that can be treated
separately \cite{parnas:71, parnas:72}. The intent with this is
to simplify the design, maintenance, and evolution of the program, 
allowing to focus
on the level of abstraction \cite{liskov:72, liskov:87}, and particular 
part of the program desired.

A software measurement program can be comparable to
a computer program in that both comprise processes, that although vary
on their nature and purpose, are essentially aimed to be executed performing
certain required tasks. 

In this paper, it is proposed to think of a measurement program as being
similar to a computer program, aiming to produce specifications
for measurement programs that are as precise and unambiguous as
possible. To this end, an specific understanding about a measurement
program and its components is established. On top of this view, a
methodology for designing software measurement programs is delineated,
in which a program is partitioned according to the concerns that motivated its
constructions in the first place, similarly as it happens in the development of computer
programs.

The necessary background is provided first, with all notions that are used to 
devise the design methodology. Then, the technique employed to separate the 
concerns of the program into modules (i.e. modularization) is explained in the context of software
measurement programs, specifically for the view considered in this paper. Subsequently,
the methodology is presented outlining some relevant steps that can be followed for the design
of a measurement program, and then, some recommendations than can lead
to a good design are proposed. 
Thereafter, an example is outlined to help picturing the methodology presented, being applied in a real
world problem. The final part of the paper talks about benefits and limitations of the methodology,
as well as the research paths that could be of interest after this work. 

\section{Background}

In order to specify a measurement program unambiguously,
it is necessary to establish a precise understanding at its respect first.
This, in no way attempts to contradict the current view found in
the literature (e.g. \cite{berry:00, fuggetta:98}), instead, it
helps to delineate programs that can be subject to a truly engineering
treatment.

A software measurement program shall be understood as a set of 
measurement tasks that are defined
and organized to satisfy a given set of measurement goals that lie 
under the concerns of a software organization.
The specification concerning the way the tasks are put together to 
achieve all measurement goals
needs to be clear and precise, so that the program can easily be analyzed, 
executed, maintained, grown, and modified.
Hence, it should define the structure of the program, 
the data and control flow that the program follows, and
it should allows to proof the completeness and correctness of the program.

The structure of a measurement program corresponds to the parts that compose
 the program together with the links that
exist between them. As a measurement program is to be seen as operational in nature, 
each of its parts should be considered to be operational as well. 
In other words,
just as a program entails a measuring activity, each of its parts
should also correspond to a simpler, yet complete, measuring activity.

Each of the measuring activities within a measurement program, regardless if it corresponds
to the whole program or to one of its parts, it generates results that can be
output, if necessary, in order to be used by other parts, or 
as the final outcome of the program.
Such results are, indeed, measurements that correspond to the data flow of the program. 
Correspondingly, the control flow 
of the program details the course of the measurement tasks within the program.
Both, data and control flow, are important in that based on them it is possible to
prove the completeness and correctness of the program.

The measurement tasks of a software measurement program are to be performed 
over specific targets that will be referred 
as objects of measurement, all of which are considered to be related to the 
development of software at some point. In this paper five kind of such objects 
are considered: products, processes, resources, organizations,
and miscellaneous (combination of any of the first four). 
These types of objects of measurement have been established given
that based on them it is possible to express any of the most common
 targets found in software measurement.
  A particular distinction is made between objects
  depending on whether they include an organization on their definition or not. 
  If the type of an object of measurement corresponds to organization,
  or to a combination that includes such type, it is said to be an
  organizational object of measurement. In other cases, the object is denominated
  as a non-organizational object of measurement.

The methodology presented adopts a goal-oriented perspective
 for the design of measurement programs, as could be seen already
from the intuitive definition of a program that was provided before.
In particular, the basic configuration of the GQM approach, as
was originally proposed by Basili and Weiss \cite{basili-weiss:84},
is used as baseline for the elicitation of the measurement goals.

Let us now clarify some terminology that will be used 
extensively throughout the paper, due to its importance 
within a goal-oriented approach for software measurement.

A measurement goal corresponds to a high-level description
of the ultimate intention of a measurement endeavor.
Hence, it should be regarded as one of the most fundamental parts of any such
endeavor and, consequently, of any measurement program as well.

The actual measurement task is driven by the metrics.
A metric defines the particular kind of measurements
that are expected to be produced by a measurement task
 
An organizational goal corresponds to the aim that the interested organization has behind 
a given measurement goal or set of measurement goals. In other words, 
it relates to the purpose 
of a measurement endeavor from the point of view of the organization.

Any measurement endeavor will define metrics, since, as explained before, 
they define
what is to be measured and how this will be done. 
Moreover it seems natural to think that measurement goals will always appear
in the process for measuring any object of measurement in a given software,
based on its fundamental importance on the definition of the endeavor.
Nonetheless, organizational goals could mistakenly be neglected or considered 
not important part of a measurement task, given that they
do not relate or derive directly into actions, and are mostly implicit
part of the elicitation of measurement goals

The view taken
in the methodology presented here is that a measurement endeavor can be based only
 in the metrics that were
derived from measurement goals, and organizational goals can be left out of consideration.
However, in order to have a measurement program
and not a simple measurement endeavor, 
additionally to the features stated in the intuitive definition of a program, given before, 
all measurement goals have to be justified or explained away in their purpose towards
 the organization for which they are being defined, or, 
in simpler terms, it is necessary to specify all organizational goals from which the measurement 
goals of the program derive,
in order to call such endeavor a software measurement program. 

\section{Modularity}

The methodology proposed is based on the application of modularization
for the design of measurement programs, in a similar way as it has been 
defined and
successfully applied for designing software systems.
The general idea is to divide the whole 
set of concerns of a measurement program into partitions or modules
 that interact with each other in order
to achieve the goals of the program. 
Each module takes care of its own tasks while interacting with other modules 
in the program. This happens either receiving measurement tasks that,
although do not lie directly under its concerns, are necessary 
for performing its tasks, or producing measurements that 
are needed by other modules.

If one considers, for example, a measurement program followed in a
 company that aims to measure some features of
an application software (e.g. its size and performance). Then, a possible separation 
of concerns could be:
\begin{itemize}
 \item \textit{module 1}: aiming to produce measurements concerning the user interface of the application
 \item \textit{module 2}: aiming to produce measurements concerning the logic of the application
\end{itemize}
Such separation, although trivial, allows, if necessary, to treat each module separately, focusing on 
the specific target assigned. Moreover, reuse of parts that compose the program become apparent. 
For instance,
\textit{module 1} can be reused directly in other measurement program performed by the same company
involving, probably, different aims while still requiring
 to measure the same from the user interface of
the application software. It can also be adapted and reused if, perhaps, a different user interface
 is to be measured,
or it can serve as the basis for building new modules inspired on a module that has been used before.

As the methodology proposed relies heavily in modularization, it is necessary first to place
this technique in context. Thus, concepts about measurement modules and connections
between such modules are given next, more from an intuitive perspective than
attempting to state formal definitions. After that, the methodology is presented explaining
how a program can be designed in a modular fashion based on the defined concepts. 

\subsection{Measurement modules}
A measurement module shall be understood as a separation of concerns within a given
measurement program, that corresponds to the execution of part of the whole
measurement endeavor. This separation is done hiding all the complexity related
to the processes of measurement while exposing only what is needed by
the other parts of the program. 

In the context of software measurement, it should not be expected for every
measurement module to be a measurement program itself, since as it will be clear later,
a module might depend on others to justify its purpose, thus failing the
requirement, mentioned in the Background section, for a measurement endeavor to
be considered as a program. Nonetheless, it is necessary for all
measurement modules to be defined as a complete measurement entity, in such a way that
one module
corresponds to, at least,
one fully constituted measurement activity. This means that, regardless of its
simplicity, a module should encompass all what is necessary to measure something within the program.

Notice that without following the requirement given above, one could not talk about
measurement modules. Instead, one would get divisions that are unable to work alone, 
thus not complying with the way modularization has been successfully used in 
software development \cite{liskov:87}. 
In what follows, only 
measurement modules will be considered,
therefore, for convenience, they will be referred simply 
as modules.

A module is characterized by three parts: interface, specification, and implementation.
The implementation corresponds to what was mentioned before as the complexity of
the module, comprising all processes it needs to perform in order to deal with 
the part of the program to which it has been assigned.
 This part is not visible to other modules and, in fact, once
a module is defined, all its implementation details are of no interest 
from the design of the measurement program point of view, as long as the module
works as expected. Thus, the focus concerning the design should be on the other parts of a module,
which are presented next.

The definition of a module is given by its specification which contains the description
of all the goals that lie under the concerns of the module. From the design of the measurement
program perspective, the specification is the most important aspect of each module,
since it gives enough information to proceed with the design.
Nevertheless, towards the stage in which modules define their implementation, 
it is necessary to settle down, precisely, what will be measured and what will be the flow of measurement information
within the program. At this respect, some modules will be expected to output their results, for other
modules use, as well as receive measurements, as inputs for its implementation. It is the
interface of each module the one that defines the kind of measurement results and inputs, if any,
that the module will produce and receive.

It is important to mention that, in general, one can loosely say that the specification of a module includes its interface,
since that still maintains the essential idea of the construction of the module, namely, to hide
its implementation from the other modules. The distinction between specification and interface is made
 in the methodology, however, because it makes explicit the possibility of designing, from a high-level perspective,
 all the measurement program, including its modules and their interactions, based only on its organizational and measurement goals,
  holding the definitions 
 of metrics to a later stage of the design.

As explained in the Background section, a measurement program should be concerned
with two types of goals: one, directly related to measurement and, the other, of organizational nature.
Measurement goals dealing with what is expected to be measured by the program, and organizational
goals relating to the purpose that the software organization has with regard to its aims of measurement. 
To cope with this, two types of modules are proposed: one, that only has measurement
goals, called regular module, and, a second type, called organizational module, that has both
organizational and measurement goals.

For the case of a regular module, the specification defines its measurement goals that, in turn,
will be used to derive the metrics that conform its interface. An organizational module,
on the other hand, has, in its specification, organizational and measurement goals according to
its concerns, which imply further differences. First, for an organizational module it might
be possible that no measurement results need to be output, given that, perhaps, it represents one 
of the highest levels on the hierarchy within the organization of interest for the program. Additionally,
the organizational goals of an organizational module will primarily
 influence its implementation that, in turn, as it will be seen later, will influence the definition of
 other modules that the first module will use to achieve its goals.
 
 The process for deriving measurement goals from organizational goals, and, correspondingly,
metrics from measurement goals, follows the GQM approach on its basic configuration as 
mentioned in the Background section.
Thus, for any kind of module (regular or organizational), one has to take the measurement goals defined on its 
specification, and, following GQM, define questions that derive into the right metrics
which constitute the interface of the module. Measurement goals, in turn, are elicited from
organizational goals that come from other modules.
Further explanation regarding measurement and organizational goals will be given when connections between
modules will be presented.

At the beginning of this subsection it was stated that a module is to be seen as a measurement entity. 
Furthermore, in the Background section it was mentioned that within a measurement program the possible 
objects of measurement correspond to products, processes, resources, organizations, and miscellaneous.
 Thus, the possible concerns of a module will encompass one or many of such objects.
 
 Also in the Background section, we established, out of the five kinds of
 objects of measurement, a further distinction, with two types of measurement objects, 
 organizational and non-organizational, depending on whether they include or not an organization in their definition.
 A similar criteria applies for the definition of modules, where the type of objects of measurement
 enclosed by a module as part of its concerns, tell us the type of module we have to define.
 Thus, if the concerns of a module involve non-organizational objects, we should define, as part of our design, a module
 that we shall call regular. On the other hand, when a separation of concerns, within
 the design of a measurement program, involves organizational objects, it implies that the organization
 or organizations in question, together with their organizational goals, are of interest for the measurement
 program, and then, we should define what we shall call an organizational module.

Given that the overall goal of this paper is to present a methodology for the design of measurement
programs, it is necessary to begin introducing, from now, two essential devices of the methodology that
will help both in its further exposition, and on its correct application afterwards.
The first device is a module diagram that is used to depict all modules of the program together with
the connections that exist between them. 

\begin{figure}[!t]%
\centering
\subfloat[][]{\includegraphics[width=0.12\textwidth]{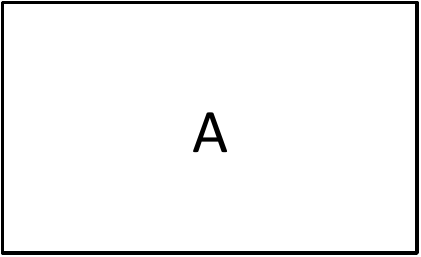}}%
\hfil
\subfloat[][]{\includegraphics[width=0.12\textwidth]{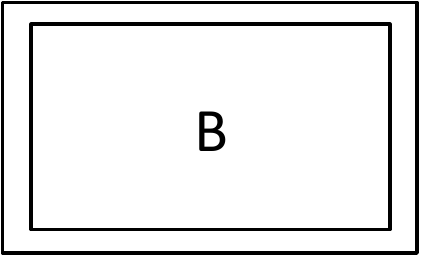}}
\caption{Types of modules in a modular diagram: {(a)~regular module} and {(b) organizational module}.}%
\label{fig1}%
\end{figure}

A module is represented by a box with its identifier appearing
in the middle. Additionally, the diagram distinguishes regular modules from organizational modules, with
a single-lined border box for the first type (Fig.~\ref{fig1}~a), and a double-lined border box for 
the second (Fig.~\ref{fig1}~b).
The extra border
should be understood as an internal layer. 
This, reflects the idea that an organizational module should follow for its construction, in
which the organizational goals that the module defines as part of its specification, are meant
to influence its own implementation, thus, their influence goes towards the internal part of the module.
 
The second essential device of the methodology relates to the unambiguous specification of the structure and data
flow of the measurement program. Based on mathematical notation, the design
of the program is precisely specified by providing all modules of the program, distinguishing between their types,
and establishing the connections that exist between them. More importantly, the notation
introduced will serve to specify relations, all along the program, between organizational goals, measurement goals,
and metrics. 

Building a precise specification of the measurement program allows to check for its completeness and
 high-level correctness.
Intuitively, a measurement program 
shall be seen as complete once it includes all necessary measurement goals, each of which has been derived to
corresponding metrics,
and from respective organizational goals. As for high-level correctness, it implies that all derivations defined in the program,
together with its structure and data flow, make sense towards achieving all goals.

To begin with the exposition of the proposed notation, let first a regular module be denoted by its identifier, same as it is
used in the module diagram. In the case of organizational modules they are denoted by an overline on top of the identifier. 
Thus, for example,
modules from Fig.~\ref{fig1} are denoted as $A$ and $\bar{B}$, respectively. The methodology do not impose any preferences
for the selection of identifiers, though, for convenience, only single capital letters of the English alphabet are used in this paper.

Organizational goals, measurement goals, and metrics are, in turn, represented as sets. Let then,
$\Gamma_{\bar{X}}$ denote the set of organizational goals of a module $\bar{X}$ and 
${\gamma_i \in \Gamma_{\bar{X}}}$ be the $i$th element of the set $\Gamma_{\bar{X}}$. Likewise,
let $G_{\bar{X}}$ denote the set of measurement goals of $\bar{X}$ and ${g_i \in G_{\bar{X}}}$ be the
$i$th element of $G_{\bar{X}}$. Furthermore, let consider $M_{\bar{X}}$ to denote the set of metrics the module
$\bar{X}$ is expected to produce, with ${m_i \in M_{\bar{X}}}$ being the $i$th element of $M_{\bar{X}}$. Thus,
 modules can be described as a collection of all its sets. In Fig.~\ref{fig1}, for example, modules are described as 
${A=\{G_A, M_A\}}$ and ${\bar{B}=\{\Gamma_{\bar{B}}, G_{\bar{B}}, M_{\bar{B}}\}}$, where in order to consider them well defined
none of their respective sets can be empty. This is important, since a complete design of a measurement program needs all modules to
be well defined.

Based on the notation given so far, it is possible to express all possible relations for any given measurement program. 
At the module level, however, only two relations are relevant to be considered, which involve only elements within the 
specification of a module. The first relation, valid for both regular and organizational modules, it is between the metrics of 
the module and its measurement goals, specifying from which goal(s) each metric is derived. 
On the other hand, 
since an organizational module defines organizational goals to govern its implementation, towards accomplishing 
its measurement goals, it is important to specify the pairwise relation between the two kinds of goals
defined within the module.

A relation is expressed via a notation that represents all corresponding dependencies as a set of ordered pairs of the form
$(a,b)$, with $b$ being dependent on $a$ in terms of execution (i.e. to achieve $b$ you need to do $a$ first), 
and $a$ being dependent on $b$ in terms of conception (i.e. $a$ is derived from the need to achieve $b$).
Let us consider, for instance, an organizational module $\bar{Y}$ with ${G_{\bar{Y}} = \{ g_1, g_2, g_3 \}}$ , 
${\Gamma_{\bar{Y}} = \{ \gamma_1, \gamma_2 \}}$, 
and ${M_{\bar{Y}} = \{ m_1, m_2 \}}$, where $\gamma_2$ was established to contribute towards accomplishing both $g_1$ and $g_2$,
 and these two measurement goals, in turn, were derived into the metric $m_1$. Additionally, $\gamma_1$ 
 was defined based on the need to accomplish
 measurement goal $g_3$, that was derived into $m_2$, correspondingly. Thus, the relations of module $\bar{Y}$ are expressed as
 \begin{flalign*}
   ~~~&G(\Gamma_{\bar{Y}}) = \{ (\gamma_2, g_1), (\gamma_2, g_2), (\gamma_1, g_3) \} \text{  and} & \\
   &G(M_{\bar{Y}}) = \{ (m_1, g_1), (m_1, g_2), (m_2, g_3) \} &
 \end{flalign*}
Now, all elements required to specify a module are in place. The complete specification of a module, including
its interface, is given as a compound of its {inputs-outputs}, goals, and corresponding relations. As an example, let us take the module
$\bar{Y}$, described above, and, for the purpose of illustrate all possible components of a complete module specification, let
us consider as the inputs of $\bar{Y}$, the metrics $m_a$, $m_b$, and $m_c$ which were defined by some other module(s) 
used by $\bar{Y}$. Therefore, the full specification of $\bar{Y}$ is
 \begin{flalign*}
  ~~~\text{\textit{module }}&\bar{Y} \coloneqq &\\
   &\text{\textit{Inputs}} = \{ m_a, m_b, m_c \} &\\
   &\text{\textit{Outputs}} = \{ m_1, m_2  \} &\\
   &\Gamma_{\bar{Y}} = \{ \gamma_1, \gamma_2 \} &\\
    &G_{\bar{Y}} = \{ g_1, g_2, g_3 \}& \\
    &G(\Gamma_{\bar{Y}}) = \{ (\gamma_2, g_1), (\gamma_2, g_2), (\gamma_1, g_3) \}& \\
    &G(M_{\bar{Y}}) = \{ (m_1,g_1), (m_1,g_2), (m_2, g_3) \}.&
 \end{flalign*}
From above, it is important to say that the interface of the module corresponds to both the set \textit{Inputs} and
the set \textit{Outputs}, where the first set
contains the metrics the module needs for its implementation,
and the second one contains 
the metrics that the module produce for other modules use. Therefore, these two sets might not be present 
in early stages of the design, or if the program
is being analyzed from high level perspective, given that in such cases only goals are relevant. 

As can be seen, no description of the relation between the measurement goals of the module and
the organizational goals from which they derive, appears in the specification of a module. This is because, although, such relation
is of interest for the program in order to prove its completeness, modules do not need to know from which organizational
goals their measurement goals derive. It should be clear that a module is an independent piece of the measurement program,
and only interact with other modules by exchanging measurements that are based on the metrics that it has defined on its interface. 
In the next subsection,
interactions between modules are clarified, with the introduction of the concept of connections.
 
 \subsection{Connections}
 
\begin{figure}
\centering
\includegraphics[width=0.30\textwidth]{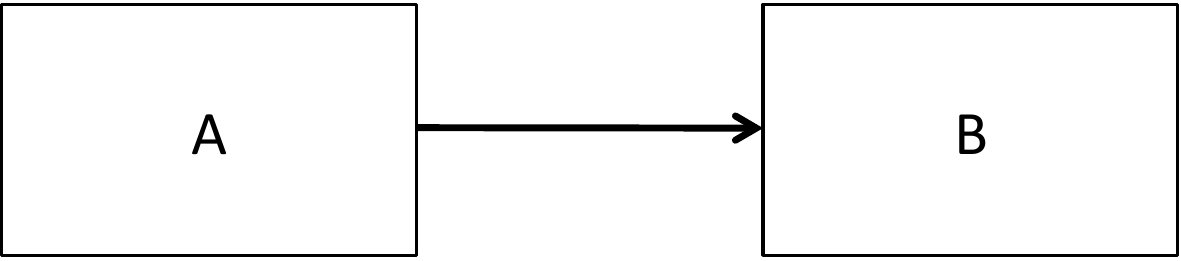}
\caption{Two modules connected ($A$ using $B$).}
\label{simple2}
\end{figure}

 In order to represent the interaction between modules, the methodology presented defines connections. 
 A connection indicates that a module
 is either using or being used by another module. The module diagram depicts a connection as an arrow, starting on the using module
 and ending in the module that is being used. This, in turn, is denoted by an ordered pair of modules as $\langle X, Y \rangle$, where the module
 $X$ is said to be using the module $Y$. For instance, Fig.~\ref{simple2} shows the diagram of a connection between modules $A$ and $B$, 
 where module $A$ is using module $B$, denoted as $\langle A, B \rangle$.
 
 From the perspective of a module, being connected implies that it will either output measurements or receive measurements
 that are needed for its implementation. Thus, connections, as presented, follow the view of Parnas \cite{parnas:71} 
 that considered them
 to be the assumptions that modules make about each other. With this in had, it should be clear that a module is interested on its connections only
 with regard to its interface, which in fact only cares for the metrics that are input or output, paying no attention to the modules 
 that use its outputs or to those from which its inputs come from. 
 
 It is when looking at connections from the perspective of the specification of a measurement program that they become 
 relevant.  
 From one side, connections complement the specification of modules in allowing to follow the data 
 flow of the program. More
 importantly, the structure of the program, as a whole, is based on the connections that, in turn, allow to keep track of the relations between 
 measurement goals all along the program, and between measurement goals and the organizational goals from which they derive.

In order to specify a connection, it is necessary to give the relation that exists between the goals involved, 
where depending on the type of modules,
the relation will focus on different kind of goals. Hence, if the connection goes from a regular module to a module of any of the two
possible types, the relation of interest should give the correspondence between the measurement goals on the first module and the 
measurement goals in the second module, thus, indicating how the second module is being used towards achieving the goals of
the first one. In contrast, if the connection originates in an organizational module, 
regardless of the type of the module in the other end, 
it has to specify the relation between the organizational goals of the using module 
and the measurement goals of the used module. 

It should be clear, from above, that in a connection we care about what is in direct influence of the
implementation of the using module, since it is its implementation the one that directly needs from the other module.
Therefore, when the using module is organizational, its organizational goals are considered for the relation, given that their
influence is more direct over the implementation of the module than the one of the measurement goals. In contrast, if 
the using module is regular, the goals that directly influence the implementation are its measurement goals, which are then
taken for the relation. On the other hand, concerning the used module,
 we care about what is in direct influence of its output interface, since
the using module will only use measurements from the output of the used module. Thus, regardless of the type of the
used module, only its measurement goals will be considered for the relation given in the specification of the
corresponding connection.

To illustrate the different implications that a connection may have, depending 
of the type of the modules involved, next are presented the four possible setups 
for a connection between two modules, classified into two groups according to
their similarities:
\begin{enumerate}[i.]
  \item $\langle X, Y \rangle$ or $\langle X, \bar{Y} \rangle$: For this kind of connections, the specification needs to give the
  relation between measurement goals in the using module and measurement goals in the used module. This, for each of these connections, is
  $G(\langle X, Y \rangle)$, with ordered pairs of the form $(g^{Y}, g^{X})$, and $G(\langle X, \bar{Y} \rangle)$, with ordered pairs of 
  the form $(g^{\bar{Y}}, g^{X})$, respectively. A superindex in the notation indicates the
 module to which the goal belongs.
 It is relevant to note that a connection that follow any of these setups, 
 cannot, by itself alone, give shape to a measurement program. 
 This is because, in such case,
 the measurement goals would not have organizational goals that explain their purpose, failing to conform with the 
 view established in the Background section about measurement programs.
 \item $\langle \bar{X}, Y \rangle$ or $\langle \bar{X}, \bar{Y} \rangle$: The specification of this kind of connections has to provide
 the relation that explains to which organizational goal in the using module is aimed to contribute a measurement goal in the used module.
 Hence, for the first connection the relation is $\Gamma(\langle \bar{X}, Y \rangle)$, with ordered pairs of the form $(g^{Y}, \gamma^{\bar{X}})$.
 Likewise, the relation to be specified for the second connection is $\Gamma(\langle \bar{X}, \bar{Y} \rangle)$, with ordered pairs
 as $(g^{\bar{Y}}, \gamma^{\bar{X}})$. Differently from the two previous setups, a connection of this kind could constitute, in itself, 
 a measurement program. For instance, if
 $\bar{X}$ would correspond to a module related to the highest level of the organization of interest for the measurement program, 
  with its interface containing only organizational goals that do not need to be explained in their purpose. Then, the two modules,
  in any of the configurations of this case,
  could be seen as a measurement program.
 
\end{enumerate}

It is important to mention that any of the cases above, regardless if itself could constitute
a measurement program or not, it can be the starting point of one. This implies that,
similarly to software development, one could take just one small part of a program, corresponding
to few modules, build the internals of each module (i.e. its implementation), to then test that part of the program and check if
the design is going as expected. 

\section{Designing Measurement Programs}

Based on the concepts described in the previous section, 
it is possible to design a measurement program in a modular fashion, where
one can separate the program not only into modules, but also, 
into subprograms that correspond to fully constituted programs that lie within
a larger program. 
In like manner, one can easily, if necessary,
 focus on portions of the program irrespectively if they 
constitute a program on their own right or not. 

For any of such cases, there will be a cleaner and more controlled implementation of the metrics,
 since all what is needed
is to follow the specifications given by the design. 
Likewise, reading, understanding, and modifying the program become easier. 
For instance, it is possible to have, very fast, a high-level view of the measurement program, 
which gives the possibility of deploying
it, from pretty early stages, and test it in order to continue with the design in an incremental fashion. 
Moreover, if one considers, for example, the case in which
at the execution of the program, it is found that the specification overlooked some metrics 
that are needed for some particular processes, it can be fairly simple to evolve 
the design, adding the functionality needed, by incorporating new module(s) that provide the missing metrics.

To illustrate the design of a measurement program from the perspective entailed by the 
methodology proposed, let us briefly examine the incremental design of a program
initialized as ${\mathcal{P} = \{ \langle \bar{A}, B \rangle \}}$, where the notation for the
specification of a program is introduced, which shows all connections and modules forming the program. 
The full specification of a program requires to
specify each module and connection as defined in the previous section. However, as our
interest will now only focus on the structure of the program, details such as goals and metrics will
not be given.

Let then, suppose that it has been found that the organization which goals are encompassed by $\bar{A}$, 
is not directly in charge of dealing with
the objects of measurement that $B$ has under its concerns. 
For this case, the designer may opt to preserve 
the regular module originally defined, or to make the organization in charge of the 
objects explicit for the measurement program. The second option is attained by 
including the organizational goals needed at that level, in order to achieve the measurement goals
 formerly established for $B$, which is done 
exchanging the regular module by an organizational one. 
In this case we decide for the second option that, as it will be clear later, would
be preferable in looking for a good design. Thus, so far our measurement program is specified
 as ${\mathcal{P} = \{ \langle \bar{A}, \bar{B} \rangle \}}$.

Given that module $\bar{B}$ has only changed from being regular to being organizational
 for the purpose of making explicit the organization in charge of its measurement tasks, its
 implementation should be considered to remain the same, since the outputs 
 expected from the module, now that is organizational, should be the same as when it was regular.  
As for $\bar{A}$, nothing has changed and its input requirements remain the same.
Here then, we assume that the designer decides that $\bar{B}$ needs another module
 in order to deal, specifically, with the measurement of
the non-organizational objects of measurement that it has under its concerns. For this purpose,
 a new module $C$ is defined, being connected to $\bar{B}$. 
 The decision of the designer in this case accounts for good design, specifically with respect to reusability.
Up to this point, our program is ${\mathcal{P} = \{ \langle \bar{A}, \bar{B} \rangle , \langle \bar{B}, C \rangle \}}$.

As last step of the incremental design that is being considered, let us suppose
that in the phase corresponding to building the implementation of modules,
it is found that the implementation of $\bar{B}$ requires some metrics
that are not currently provided by C.
Let us suppose also, that the metrics needed by the implementation of
$\bar{B}$ had been defined some time before our program were designed,
as part of an isolated effort within the organization, that only defined metrics
and collected the corresponding data, without following a goal-oriented approach.

We then consider that the organization in charge accepts to use the existent metrics and data,
having to reverse engineering the process in order
 to define a new module with a specification that preserves the completeness of the program. Hence, the goals of
 the new module, identified as $D$ are elicited from the existent metrics.
 
\begin{figure}
\centering
\includegraphics[width=0.25\textwidth]{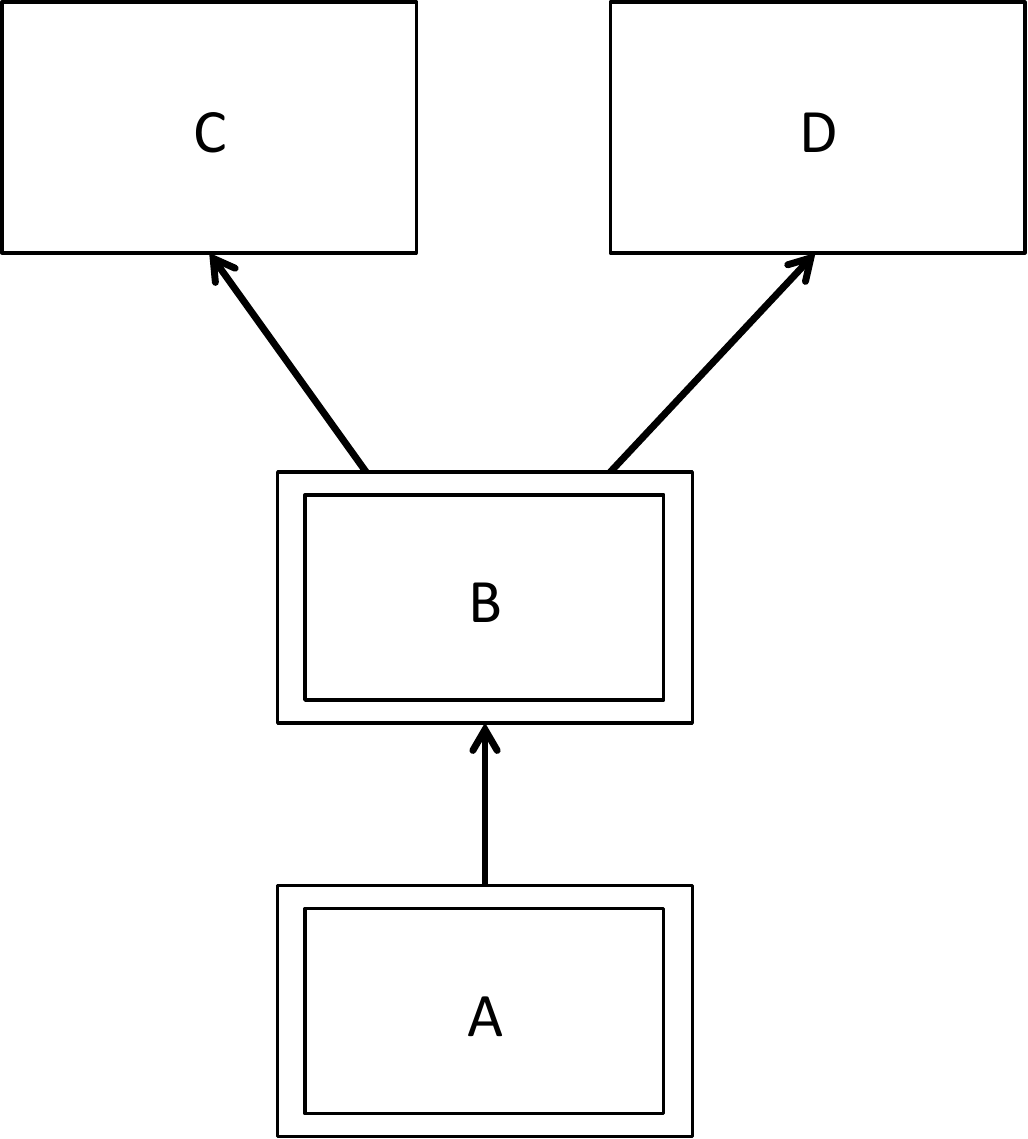}
\caption{Module diagram of a software measurement program built incrementally.}
\label{multiple4}
\end{figure}

 At last, after the inclusion of $D$, the program $\mathcal{P}$ that has been 
 built incrementally is specified as ${\mathcal{P} = \{ \langle \bar{A}, \bar{B} \rangle , \langle \bar{B}, C \rangle, \langle \bar{B}, D \rangle \}}$, 
 and depicted as module diagram in Fig.~\ref{multiple4}

It can be noticed that a design based on modularization as illustrated above, has a hierarchical structure built
implicitly (in Fig.~\ref{multiple4} $\bar{A}$ lies in the first or highest level, and $C$ and $D$ in the lowest level).
 The structure, as could be seen from the previous section, does not impose any precedence restriction concerning
one-to-one connections. Instead, it is when looking at the program as a whole that the hierarchy serves its purpose, being an
aid for ensuring completeness and good design of the program. 

It is important to point out that the relevant levels of the program, when considering its hierarchy, 
are only the highest and lowest levels, specifically if we find organizational modules. 
From one side, only modules of the type organizational can occupy the highest level in the hierarchy,
since otherwise some measurement goals will remain without explaining the organizational goals from which they derive, 
making the program incomplete.
On the other hand, in what respects to the lowest level, from good design perspective, as will be explained later, 
it is in general not a good idea
to have organizational modules at the last level of the hierarchy. 

\subsection{Good Design}

The main motivation of presenting a design methodology for measurement programs is to provide a 
way to design programs that can grow, evolve, be maintained and reused, 
in different circumstances either by parts or as a whole.
Up to this point, the necessary elements to design a measurement program based on modularization 
have been established, however, it
has not been said how a good design can be achieved. It is clear that the qualification about
how good is some thing can be rather tricky, thus, rather than giving a precise way to
determine such quality on a design, some pointers that can be found important in aiming for a good design are presented next:

\begin{itemize}
 \item Modules should be defined following, first, the concerns of the measurement program, and then, considering the structure,
 the organization, and the features of the objects of measurement. In other words, it is important to keep in mind that although 
 the measurement
 program will follow a modular structure, and the objects of measurement might be also defined based on modularity, 
 their concerns are fundamentally different. Thus, it might happen, and be a good design for a measurement program,
 that its structure follow some parts of the structure of the objects of measurement, as long as the actual concerns of
 the measurement program are the ones that
 lead the design. 
 \item At all times, any partition should respond
 to the need of organizing the program towards the achievement of its measurement goals.
 \item All metrics should be made explicit at the specification level. 
 This implies that no metric will remain hidden in the implementation
 of a module, not even if it represents an intermediary step towards the actual concerns.
 \item The inputs and outputs of a given module should be limited in number. It is not possible
 to establish a precise number for this, or even to try to propose some thresholds, since
 this will depend on the particular program in hand, and on the overall design that is being done. Nonetheless,
 it is important to maintain modules in a good balance of complexity, given that one of the main ideas
 of applying modularization for the design of measurement programs is to make them, in general, easily manageable.
 Thus, for example, if a module is requiring many metrics for its implementation, or it has to produce a large amount
 of metrics for other modules use, it is an indicator that the module in question is probably too complex and should be 
 split into as many modules as needed, in order to make each of them easily manageable. 
 \item It is
 important to make sure that a module does not have to deal alone with a large number of measurement concerns, or that
 those concerns do not represent, in themselves, very complex endeavors. A module should, then, be designed maintaining a
 balance in the complexity of the processes of its implementation, and on the measurements that exchanges with the 
 other parts of the program. To this end, it is necessary to avoid the creation of modules that encompass many different kinds
 of objects of measurement, because, as it has been said before in this paper, although the concerns of a module might
 be defined to measure a compound of more than one kind of object, if such combination in itself
 is complex, it would make the module hard to handle from its definition.
 This is especially true when organizational concerns are in play within a module, given that its nature differs greatly
 from non-organizational modules.
 \item Organizations that are directly related to an object of measurement should be made explicit
 in the program. 
 The reason behind this point is that the organization that is in direct relation to an object
 will better know how to deal with it, regardless if its relation corresponds to measuring activities or not. 

 \end{itemize}

The list given above is in no way extensive or very detailed. The
 most important
idea to keep in mind in order to attain a good design of a measurement program, 
is to look for obtaining a program that can be better understood, reused, maintained,
and modified. All these are indeed features
that should be aimed in any good engineering design.

\subsection{Design Steps}

After having provided a general idea of how to design a program based on the methodology proposed, and
having established some important recommendations towards the achievement of a good design, understood
under the criteria presented. It is possible to give an outline of the steps that can be taken 
following the methodology, in order to produce a good design of a measurement program.
The following sequence of steps has to be considered as a main course of execution, where
iteration can happen at any point, if necessary:

\begin{enumerate}
 \item The first step is to give a preliminary list with all organizational and measurement goals that
 can be elicited easily (i.e. goals that are clear or obvious) from the initial definition of the measurement program.
  At this point the most important thing is
 to gather the measurement goals that are essential for the definition of the program, at least at first glance.
 \item The goals gathered in the previous step have to be, depending on its nature, 
  derived into measurement goals or explained away based on 
 the organizational goals from which they derive. As mentioned previously in this paper, 
 GQM is recommended for a correct elicitation of goals. 
 It is important to mention that the first two steps should not aimed to 
 obtain a detailed, complete, and definitive list of the goals of the measurement program, 
 since it is expected that more goals
 will become apparent when modularization begins to be applied.
 \item Considering the goals of the program that have been found so far, modules are defined by establishing how the measurement 
 concerns can be better divided. Additionally, connections that follow clearly from the defined modules, and from the relation between
 organizational and measurement goals have to be established without giving its complete specification yet, i.e., no data flow
 is specified at this point. Such connections have to be considered as preliminary, since by not being properly specified
 they cannot serve the purpose of a connection yet.
 \item This step represents a combination of actions that rather than being performed sequentially, are expected
 to support each other on their progress. 
 Splitting and merging over the previously defined modules happens in this step, together with the consequent
 initial definition of high-level data flow, i.e., specification of the connections between modules. 
 It is at this stage when both goal elicitation (based on GQM) and modularization need to be combined
 in order to decide for the best way to divide the program into measurement concerns, to attain a good design of the
 measurement program.
 Thus, for instance, in this step the designer would be taking the decision that was discussed 
 at the beginning of this section about making a regular module become
 organizational ($B$ for $\bar{B}$ in the example), given that the consideration of the organization 
 benefits a better design. Likewise, good part of the connections
 of the program are specified
 as part of the modularization process, yielding already an structure that is more than halfway to be completed.
 \item All the connections of the program are settled down in this step. In addition,
 at this point every module that needs to be added to the program to diminish the complexity of modules, 
 especially in the case of 
 organizational modules, has to be defined and connected. As an example, 
 it would be here where the designer of the program outlined
 at the beginning of the section, 
 would define $C$ and $D$ in order to help the module $\bar{B}$.
 Additionally, it is important to say that although the definition of metrics is being left for a later stage of the design, 
 if the definition of some metrics
 can be helpful to complete the definition of the connections of the program, they should be defined already here.
 \item At this step a hierarchy level check is conducted. This means to look for the hierarchy level 
 to which each module corresponds,
 in the way the design has been established so far, and decide on the modules that are going to be preserved 
 in the lowest and
 , especially, in the highest level of the measurement program. 
 For the case of the lowest level, one should look for organizational modules that have been defined there,
 and iterate back to step 4 in order to split all such modules leaving only regular modules in the lowest level of the program
 (following the recommendation on good design provided at this respect hereinbefore).
 The key part of the hierarchy level check focuses on the highest level,
 where all modules that are not used by others are initially considered to be in the highest level. 
 Then, from these ones, it has to be decided
 whether all of them truly should lie in the highest level of the program or not. The implication here is that all modules that 
 lie in the highest level
 \textemdash~of organizational type of course \textemdash~represent programs on their own and they could be executed 
 independently from the others. 
  The important thing to decide is if they should be taken as sub-programs
 of the measurement program that is being designed, or is a better idea to take them separately.
 Notice that, although no specific consideration about control flow of a measurement program is made by the methodology presented, 
 having sub-programs 
 underlies a more or less unified execution among all sub-programs. On the other hand, breaking up the program
 into separate programs implies that the execution of one should not have any consideration with respect to the other ones.
 For a program as in Fig.~\ref{multiple4}, for instance, a hierarchy level check would make no difference,
  since, only regular modules lie in the lowest level of the program, and only one module lies in the highest hierarchy level.
 \item All metrics are defined in this step. This might signify a further separation of modules, which only calls for a iteration back
 to step 4.
 \item The final step of the process is to check for the completeness and high-level correctness of the measurement program. All 
 measurement goals have to be included with their corresponding derived metrics and organizational goals from which they derive. 
 Additionally, it
 has to be assessed the usefulness and practicality of the design and its parts, with respect to the
  achievement of the goals of the program.
\end{enumerate}

Certainly, in the steps above there has been no explicit consideration of cases where reverse engineering should
be the option to follow, if metrics are available from the past but they have not been produced following a goal-driven approach
(as in the example of incremental design at the beginning of the section).
Nevertheless, it should be understood that at all times the designer has to consider things that has been produced in the past, 
from measurement programs that were well designed, to data that needs to be properly formatted or put in context. 

Iteration, as mentioned before, should be part of the process for designing a measurement program, going forth and back
in the steps as much as needed. As noticed, iteration will not go back to the beginning of the design most of the times, instead,
it will go to
steps 4 or 5, where modularization is heavily employed to produce a good design.

One can see that by creating measurement programs in the way described in this paper,
they can be seen much as programs are seen in software development: pieces
that although independent of others, they can be combined with others in order to build larger programs. 
Pieces that
can evolve, be modified and expanded, in a clean and controlled manner.

\section{Example}

The initial setup of the example presented in this section is inspired by different real world projects that have
been studied at the Department of Computer Science of the University of Helsinki, on
its Software Factory \cite{abrahamsson:10, fagerholm:13}.
As the intent of the example is to illustrate some of the benefits that can be derived from applying the methodology proposed 
for the design of software measurement programs, only details that help to construct such understanding are considered.

A middle size software company aimed to improved the satisfaction of its customers. To this end, the company had
 already considered
to look for possible factors affecting customer satisfaction, with the collaboration of two internal units, Software
Development and Customer Care departments. The idea at this point was not clear about what to measure and how to do it. 
However
the company knew that some measurement activity was required in both departments, 
and that the measurements gathered on each side
should be processed to make apparent the relation between product features and the satisfaction of its customers.

From the setup described above, it seems clear that some modules can be defined. 
One organizational module $\bar{A}$, concerned with the
main goal of the company . Then, two more modules, $\bar{B}$ and $\bar{C}$, each for taking into account the organizational
goals that are to be defined with regard to the Software Development and Customer Care departments,
 towards measurement goals yet to be specified.

One more module can be defined from what is mentioned in the initial setup. 
Notice that the collaboration between the two departments
will better become a project (i.e. a miscellaneous
 object of measurement) if it is not already the case. 
 Then, a decision towards good design, would be to separate the concerns
about the project from the concerns of the organizations, promoting reusability and avoiding complex modules. 
Thus, a regular module
$D$, is defined to deal with the measurement concerns related to the collaboration project, providing
 coherent processing to the metrics
 coming from different
departments, and delivering what is needed for the organizational goal of $\bar{A}$. 
This decision will allow, as well, to include easily in the concerns of the program other departments 
that could be helpful for the collaboration project, by defining new modules and its connections to $D$.

For the case of modules $\bar{B}$ and $\bar{C}$, it is clear that they will use other modules, both from a good design perspective,
 and responding to the evolution of the design that most probably in the future will require to measure more elements. 
 Each department had already done
some preliminary work on its own, without following any specific methodology. 
Customer Care had gathered measurements corresponding to the 
satisfaction of the customers with respect to
the company in general. Software Development, in turn, had obtained measurements from two of its developed products. 

As in both cases above the measurements were obtained with no specific methodology for the derivation of metrics,
it is possible to define a module for each measurement endeavor by reverse engineering the measurement results
to obtain the corresponding measurement goals and metrics. Thus, for the concerns about measuring products, two
modules, $E$ and $F$, are defined, one for each product. Similarly, for the measurement endeavor previously followed 
by the Customer Care department, a module $G$ is defined.

\begin{figure}
\centering
\includegraphics[width=0.35\textwidth]{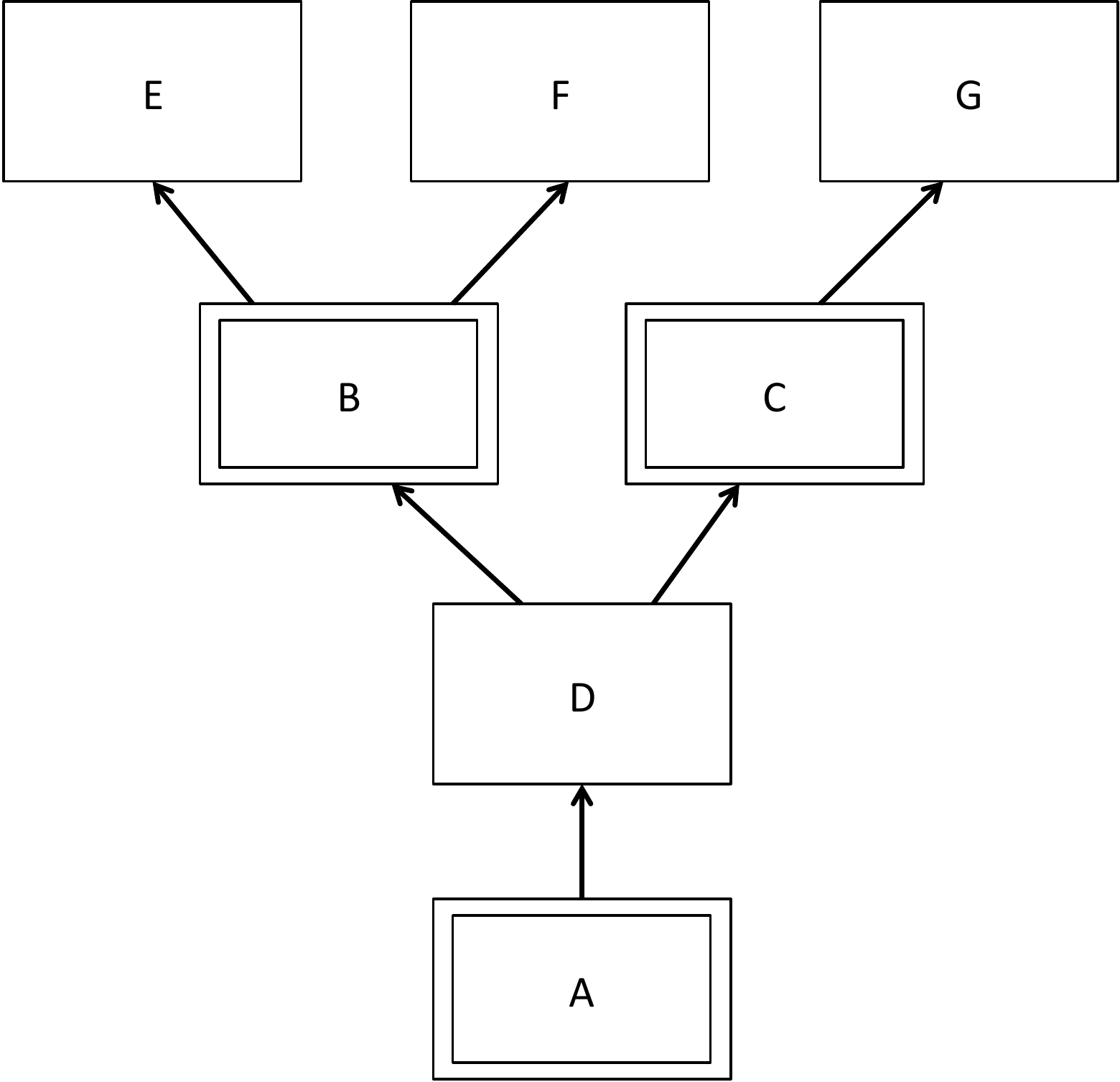}
\caption{Module diagram of an example of a software measurement program inspired by real world cases.}
\label{example}
\end{figure}

Up to this point the structure of the program stays as 
${\mathcal{P} = \{ \langle \bar{A}, D \rangle, \langle D, \bar{B} \rangle, \langle D, \bar{C} \rangle, \langle \bar{B}, E \rangle, 
\langle \bar{B}, F \rangle, \langle \bar{C}, G \rangle \}}$,
which is depicted as module diagram in Fig.~\ref{example}. Although in this example no specific goals and metrics are mentioned, it
is possible to see how the program not only is being designed with a precise structure, 
but also it is fairly simple to continue its design defining
more modules and connections, as required.

Assuming that all goals and metrics were defined, it would be possible, already at this stage,
 to test the program and see if everything goes as expected.
Moreover, the design given, allows for a smooth evolution of the program. 
From the perspective of module $\bar{B}$ one could expect, among other things, that more products will have to
be considered and probably more metrics will be needed for each of the products. 
In the first case, as more products will be considered, a good idea would be to include
new modules to introduce in the concerns of the measurement program 
the specific teams within the department on charge of each product, assuming, of course, that the organization
of the department follows such structure. In this way, organizational modules will be used by $\bar{B}$ as 
intermediaries that divide the endeavors making less complex its execution.
On the other hand, if more metrics will be needed, it might be necessary to separate the concerns 
so that each module deals with a limited amount of them. Thus, $E$ and $F$ would use
other new modules that correspond to a coherent separation of its respective measurement endeavors.

Looking at module $\bar{C}$, one could think of adding more measurement endeavors
 concerning customer satisfaction with respect to different specific
aspects of the company, such as products, support, services, etc. 
At this respect, a similar idea, as with module $\bar{B}$, could be applied.

Going a bit further, it might be necessary to acquire not only more information with respect to different aspects, 
but also to get a
whole new perspective including another department into the collaboration project. 
Thus, assuming that the organization of the company includes a Support department, as indeed is customary for
companies of this size. One could include in the concerns of the program measurement endeavors directly
related to the Support department, in which case, it would be good idea to include a 
module to be used by the collaboration project, in order to deal with such concerns.

Notice that at all times, the evolution of the design do not need very complicated modifications, 
nor require to dive into an enormous amount of information that describes the program,
in order to find where could be good to include more things to the program. Instead, 
a clear specification of the measurement program allows for localized additions and
modifications that do not affect parts that should not be concerned with the change. 
Additionally, the program has a simple yet detailed structure that, as has been seen,
can be handle from a high level perspective with no much consideration on low level details.

\section{Conclusion}

The main point of this paper has been to propose a methodology for the design of software measurement programs.
To this end, the idea of measurement programs has been defined and detailed, so that the can be specified
in a precise and effective way. The necessary components of the specification of a measurement program have
been note and explained as well.

The methodology has been built on top of the understanding given about measurement programs, 
using the technique of modularization as a vehicle for achieving a design that contemplates the
components required for the specification of a program, providing desired features for its further
modification, growth, and evolution.

Although the methodology is complete and fully applicable to real world problems,
as outlined by the example, it does not provide means to completely specify the control flow
of a measurement program.  The reason for this relies on the idea that the main goal
of the methodology is to provide coherent and precise structure to a program first, to then,
on top of this structure, establish the flows of the program easily. Therefore, the flows are considered secondary,
or at least subordinated to the structure.

Thus, in this paper the point concerning control flow is left more of less open, expecting from future 
works the establishment of tools for its precise specification. At this respect, the author of this paper 
finds himself already working on a tool similar to a programming language that, inspired by
the idea proposed by Leon J. Osterweil \cite{osterweil:87} for the specification of software processes,
will allow to specify, precisely and with no ambiguity, software measurement processes. 
This, certainly will comprise the different levels all along the program, from the processes
that occur within each module, to the processes that drive the measurement program at its highest level.


\begin{IEEEbiography}[{\includegraphics[width=1in,height=1.25in,clip,keepaspectratio]{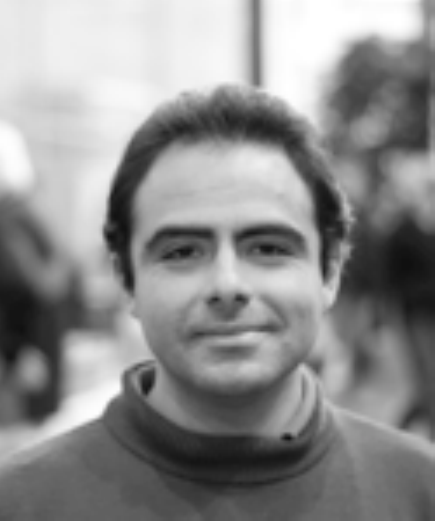}}]{Alejandro S\'anchez Guinea}
studied Computing Engineering at the Universidad Nacional Aut\'onoma de M\'exico (National Autonomous University of Mexico). 
He received a Master degree in Embedded Systems from the Institute Sup\'erieur de l'Aeronautique et de l'Espace
in Toulouse, France. He currently works at the Department of Computer Science of the University of Helsinki in Finland,
collaborating with the Software Systems Engineering Research Group. His main research interests are software
 measurement, software processes improvement, and software engineering design methodologies.
\end{IEEEbiography}


\begin{thebibliography}{10}
\providecommand{\url}[1]{#1}
\csname url@samestyle\endcsname
\providecommand{\newblock}{\relax}
\providecommand{\bibinfo}[2]{#2}
\providecommand{\BIBentrySTDinterwordspacing}{\spaceskip=0pt\relax}
\providecommand{\BIBentryALTinterwordstretchfactor}{4}
\providecommand{\BIBentryALTinterwordspacing}{\spaceskip=\fontdimen2\font plus
\BIBentryALTinterwordstretchfactor\fontdimen3\font minus
  \fontdimen4\font\relax}
\providecommand{\BIBforeignlanguage}[2]{{%
\expandafter\ifx\csname l@#1\endcsname\relax
\typeout{** WARNING: IEEEtran.bst: No hyphenation pattern has been}%
\typeout{** loaded for the language `#1'. Using the pattern for}%
\typeout{** the default language instead.}%
\else
\language=\csname l@#1\endcsname
\fi
#2}}
\providecommand{\BIBdecl}{\relax}
\BIBdecl

\bibitem{basili-rombach:88}
V.~Basili and H.~Rombach, ``The {T}ame {P}roject: {T}owards {I}mprovement-{O}riented
  Software Environments,'' \emph{IEEE Trans. Softw. Eng.},
  vol.~14, no.~6, pp. 758--773, June 1988.

\bibitem{fenton:94}
N.~Fenton, ``Software Measurement: a Necessary Scientific Basis,'' \emph{IEEE Trans. Softw. Eng.}, 
vol.~20, no.~3, pp. 199--206, Mar. 1994.

\bibitem{morisio:99}
M.~Morisio, ``Measurement Processes are Software, too,'' \emph{J. Syst. Software}, 
vol.~49, no.~1, pp. 17--31, Dec. 1999.

\bibitem{kitch:95}
B.~Kitchenham, S.~Pfleeger, and N.~Fenton, ``Towards a Framework for Software
  Measurement Validation,'' \emph{IEEE Trans. Softw. Eng.},
  vol.~21, no.~12, pp. 929--944, Dec. 1995.

\bibitem{briand:96}
L.~Briand, S.~Morasca, and V.~Basili, ``Property-Based Software Engineering
  Measurement,'' \emph{IEEE Trans. Softw. Eng.}, vol.~22,
  no.~1, pp. 68--86, Jan. 1996.

\bibitem{briand:02}
L.~Briand, S.~Morasca, and V.~Basili, ``An Operational Process for Goal-Driven Definition of Measures,''
  \emph{IEEE Trans. Softw. Eng.}, vol.~28, no.~12, pp.
  1106--1125, Dec. 2002.

\bibitem{gopal:02}
A.~Gopal, M.~Krishnan, T.~Mukhopadhyay, and D.~Goldenson, ``Measurement
  Programs in Software Development: Determinants of Success,'' \emph{IEEE Trans. Softw. Eng.}, 
  vol.~28, no.~9, pp. 863--875, Sept. 2002.

\bibitem{plfeeg:93}
S.~Pfleeger, ``Lessons Learned in Building a Corporate Metrics Program,''
  \emph{IEEE Softw.}, vol.~10, no.~3, pp. 67--74, May 1993.

\bibitem{basili-weiss:84}
V.~Basili and D.~Weiss, ``A Methodology for Collecting Valid Software
  Engineering Data,'' \emph{IEEE Trans. Softw. Eng.}, vol.
  SE-10, no.~6, pp. 728--738, Nov. 1984.

\bibitem{briand-practical:96}
L.~C. Briand, C.~M. Differding, and H.~D. Rombach, ``Practical Guidelines for
  Measurement-Based Process Improvement,'' \emph{Software Process Improvement
  and Practice}, vol.~2, no.~4, pp. 253--280, Dec. 1996.

\bibitem{basili:92}
V.~Basili, G.~Caldiera, F.~McGarry, R.~Pajerski, G.~Page, and S.~Waligora,
  ``The Software Engineering Laboratory - an Operational Software Experience
  Factory,'' \emph{Int'l Conf. Software Eng.},
  pp. 370--381, 1992 .

\bibitem{basili:93}
V.~R. Basili, ``The Experience Factory and its Relationship to other
  Improvement Paradigms,'' \emph{Proc. Fourth European Software Eng. Conf. (ESEC)}, Sept. 1993.

\bibitem{offen:97}
R.~J. Offen and R.~Jeffery, ``Establishing Software Measurement Programs,''
  \emph{IEEE Softw.}, vol.~14, no.~2, pp. 45--53, Mar. 1997.

\bibitem{becker:99}
S.~A. Becker and M.~L. Bostelman, ``Aligning Strategic and Project Measurement
  systems,'' \emph{IEEE Softw.}, vol.~16, no.~3, pp. 46--51, May 1999.

\bibitem{alejandro:01}
A.~Alejandro, ``Management Indicators Model to Evaluate Performance of IT
  Organizations,'' \emph{Portland Int'l Conf. Managem. Eng. Technol. (PICMET)}, pp.28, 2001. 

\bibitem{basili:07_1}
V.~Basili, J.~Heidrich, M.~Lindvall, J.~Munch, M.~Regardie, and A.~Trendowicz,
  ``GQM$^+$Strategies--Aligning Business Strategies with Software
  measurement,'' \emph{First Int'l Symp. Empirical Softw. Eng. Measurem. (ESEM)}, pp. 488--490, Sept. 2007.

\bibitem{basili:07}
V.~Basili, J.~Heidrich, M.~Lindvall, J.~M{\"u}nch, M.~Regardie, D.~Rombach,
  C.~Seaman, and A.~Trendowicz, ``Bridging the Gap between Business Strategy
  and Software Development,'' in \emph{Twenty Eighth Int'l Conf. Information Syst.}, pp. 1--16, 2007.

\bibitem{basili:10}
V.~R. Basili, M.~Lindvall, M.~Regardie, C.~Seaman, J.~Heidrich, J.~Munch,
  D.~Rombach, and A.~Trendowicz, ``Linking Software Development and Business
  Strategy through Measurement,'' \emph{Computer}, vol.~43, no.~4, pp. 57--65, Apr.
  2010.

\bibitem{baldwin:00}
C.~Y. Baldwin and K.~B. Clark, \emph{Design Rules: The Power of
  Modularity}. The MIT Press, 2000, vol.~1.

\bibitem{parnas:71}
D.~L. Parnas, ``Information Distribution Aspects of Design Methodology,'' \emph{Proc. of the IFIP Congress}, Aug. 1971.

\bibitem{parnas:72}
D.~L. Parnas, ``On the Criteria to be Used in Decomposing Systems into Modules,''
  \emph{Commun. ACM}, vol.~15, no.~12, pp. 1053--1058, Dec. 1972.

\bibitem{liskov:72}
B.~H. Liskov, ``A Design Methodology for Reliable Software Systems,'' 
  \emph{Proc. AFIPS}, pp. 191--199, Dec. 1972.

\bibitem{liskov:87}
B.~Liskov, ``Keynote Address-Data Hbstraction and Hierarchy,'' in \emph{ACM
  SIGPLAN Not.}, vol.~23, no.~5,
   pp. 17--34, Jan. 1987.

\bibitem{berry:00}
M.~Berry and R.~Jeffery, ``An Instrument for Assessing Software Measurement
  Programs,'' \emph{Empirical Software Engineering}, vol.~5, no.~3, pp.
  183--200, 2000.

\bibitem{fuggetta:98}
A.~Fuggetta, L.~Lavazza, S.~Morasca, S.~Cinti, G.~Oldano, and E.~Orazi,
  ``Applying GQM in an Industrial Software Factory,'' \emph{ACM Trans. Softw. Eng. and Methodology}, vol.~7, no.~4, pp. 411--448,
  Oct. 1998.

\bibitem{abrahamsson:10}
P.~Abrahamsson, P.~Kettunen, and F.~Fagerholm, ``The Set-up of a Software
  Engineering Research Infrastructure of the 2010s,'' \emph{Proc. 11th Int'l Conf. Prod. Focused Softw. (PROFES)}, 
  pp. 112--114, 2010.

\bibitem{fagerholm:13}
F.~Fagerholm, N.~Oza, and J.~M\"unch, ``A platform for Teaching Applied
  Distributed Software Development: The Ongoing Journey of the Helsinki
  Software Factory,'' \emph{Collaborative Teaching of Globally Distributed
  Software Development}, 2013.

\bibitem{osterweil:87}
L.~Osterweil, ``Software Processes are Software too,'' \emph{Proc. 9th Int'l Conf. Software Eng.}, pp. 2--13 , 1987.

\end{thebibliography}
\end{document}